\documentclass[letterpaper,12pt]{article}
\usepackage[left=2.54cm,top=2.54cm,right=2.54cm,bottom=2.54cm]{geometry}
\usepackage{makeidx}         
\usepackage{graphicx}        
\usepackage[bottom]{footmisc}
\usepackage{natbib}          
\usepackage{tipa}
\usepackage{amsmath}
\usepackage{amsfonts}
\usepackage[utf8]{inputenc}
\usepackage[T1]{fontenc}
\usepackage{color}
\usepackage{upgreek}
\usepackage{url}
\usepackage{authblk}
\usepackage[margin=1cm]{caption}

\newcommand{\nub} {{\boldsymbol{\nub}}}

\newcommand{\etab} {{\boldsymbol{\eta}}}

\newcommand{\Sset} {\mathcal{S}}
\newcommand{\intd} {\textrm{d}}
\newcommand{\phib} {\boldsymbol{\phi}}

\newcommand{\bvec} {\textbf{b}}

\newcommand{\xvec} {\textbf{x}}

\newcommand{\svec} {\textbf{s}}

\newcommand{\betab} {\boldsymbol {\beta}}

\title{A Review of Bayesian Modelling in Glaciology}
\author[1]{Giri Gopalan}
\author[2]{Andrew Zammit-Mangion}
\author[3]{Felicity McCormack}

\begin{document}

\affil[1]{Statistics Department, College of Science and Mathematics, California Polytechnic State University, San Luis Obispo, CA, USA, \texttt{ggopalan@calpoly.edu}}

\affil[2]{School of Mathematics and Applied Statistics and Securing Antarctica’s
Environmental Future, University of Wollongong, Wollongong, NSW, Australia, \texttt{azm@uow.edu.au}}

\affil[3]{School of Earth, Atmosphere \& Environment and Securing Antarctica’s
Environmental Future, Monash University, Clayton, VIC, Australia, \texttt{felicity.mccormack@monash.edu}}

\setlength{\affilsep}{2em}
\renewcommand\Authands{ and }
\date{}
\maketitle

\abstract{Bayesian methods for modelling and inference are being increasingly used in the cryospheric sciences, and glaciology in particular. Here, we present a review  of recent works in glaciology that adopt a Bayesian approach when conducting an analysis. We organise the chapter into three categories: i) Gaussian-Gaussian models, ii) Bayesian hierarchical models, and iii) Bayesian calibration approaches. In addition, we present two detailed case studies that involve the application of Bayesian hierarchical models in glaciology. The first case study is on the spatial prediction of surface mass balance across the Icelandic mountain glacier Langj\"{o}kull, and the second is on the prediction of sea-level rise contributions from the Antactcic ice sheet. This chapter is presented in such a way that it is accessible to both statisticians as well as earth scientists. This work will appear as a chapter of the forthcoming book titled \textit{Statistical Modelling Using Latent Gaussian Models - with Applications in Geophysics and Environmental Sciences}, expected to be published by Springer in 2022.}

\section{Introduction}

\begin{figure}[t!]
  \centering
    \includegraphics[width=.7\textwidth]{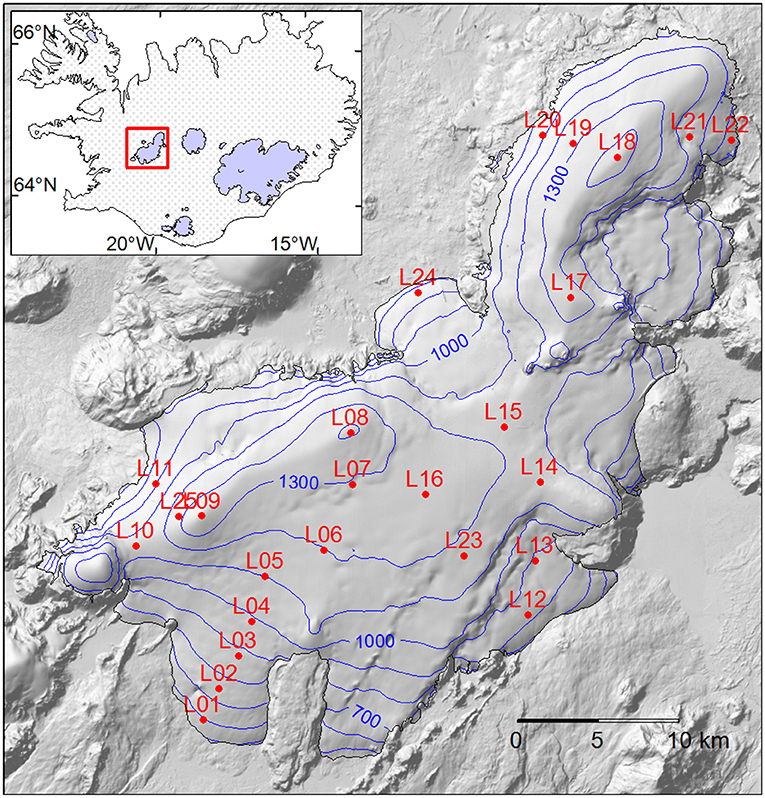}
\caption{Surface topographical map of Langj\"{o}kull from \citet{10.3389/feart.2021.610069}. The inlet map illustrates the major Icelandic glaciers, with the largest, Vatnaj\"{o}kull, to the southeast. Blue lines denote elevation contours, with contour labels denoting elevation in meters. The red dots and accompanying labels denote locations at which surface mass balance measurements are recorded twice-yearly.\label{fig:intro}} 
\end{figure}

 Ice sheets and glaciers are large bodies of ice formed  from the compaction of accumulated snow, over centuries to millennia. These bodies of ice range in size from the ice sheets of Antarctica and Greenland to comparatively smaller mountain glaciers such as those of Iceland (e.g., Vatnaj\"{o}kull and Lang\"{o}kull; see Figure~\ref{fig:intro}); about 10\% of the earth's surface is covered in ice \citep{Paterson}. Understanding the past, present, and future behaviour of glaciers and ice sheets is of major importance in our warming world because these masses of ice are the largest potential contributors to sea-level rise \citep{fox2021ocean,hock2019high}.

Glaciers are dynamic and continuously flowing in the direction of steepest ice-surface slope due to the force of gravity. This flow can be described using mathematical equations, typically partial differential equations based on Stokes’ law or some approximation (e.g., a linearisation) thereof \citep{Blatter_1995,Macayeal_1989,Pattyn_2003}. These equations relate the flow of ice to spatially-varying parameters such as ice geometry (surface slope, thickness), ice properties (temperature, microstructure), and the subglacial environment (bed substrate, hydrology). Ice flows through a combination of ice deformation, basal sliding (as meltwater acts as a lubricant for the ice to slide over the bed), and deformation of the bed.

Statistical modelling has a significant role to play in glaciology for several reasons. First, it can be used to provide probabilistic predictions (i.e., estimates with associated uncertainties) of processes of interest at unobserved locations. This is important, as glaciological data are often sparse and not evenly sampled amongst a spatial domain of interest:  in situ experiments for data collection in a remote place like Antarctica are time intensive and very expensive, while it is notoriously difficult to obtain good spatial coverage from earth-observing satellites at high latitudes. 
Second, statistical modelling allows one to infer important quantities of interest from indirect observations. This is useful because in cryospheric sciences it is often the case that collected data are only loosely related to the inferential target. For example, the satellite mission Gravity Recovery and Climate Experiment (GRACE) is able to provide a time series of mass change of an area covering part of an ice sheet. However, this observed mass change does not only include that which is due to ice processes, but also that which is due to solid-earth processes such as glacio-isostatic adjustment. Statistical methods can be used to provide probabilistic assessments on what the drivers of the observed mass change are. Third, a statistical model can be used for quickly \emph{emulating} a complex numerical model, which takes a long time to run, and also to account for uncertainty introduced by not running the full numerical model. This is relevant for studies in glaciology, as the only way to reliably project an ice sheet's behaviour into the future is via the use of an ice-sheet model such as the \textit{Ice-sheet and Sea-level System Model} \citep[ISSM, e.g., ][]{Larour_2012} or \emph{Elmer/Ice} \citep[e.g.,][]{Gagliardini_2013}, which require considerable computing resources. These ice-sheet models need to be equipped with initial and boundary conditions that need to be calibrated from data or other numerical models; statistical emulators thus offer a way forward to make the calibration procedure computationally tractable. These strengths of statistical modelling and inference have led to a vibrant and active area of research in statistical methodology for glaciology, and the widespread use of sophisticated, often Bayesian, inferential techniques for answering some of the most urgent questions relating to glaciological processes and their impact on sea-level rise.  

Glaciologists have used a variety of techniques to produce estimates of important glacial quantities from data, most notably the class of \emph{control methods} that formulate the estimation procedure as a constrained optimisation problem. There is often no unique solution to an inverse problem given a particular set of measurements, particularly when the number of unknown parameters far exceeds the number of measurements; moreover, solutions to an inverse problem may be very sensitive to small changes in measurements. One way to combat these problems in glaciology is to use regularisation via constrained optimisation, but many sources of uncertainties are not typically accounted for adequately in this way. 

Many early works employed  control methods for estimating the spatially-varying basal friction coefficient -- an unobservable spatially-varying parameter that informs the basal boundary condition or basal friction law in the momentum equations -- by minimising the mismatch between the observed and modelled surface velocity data \citep{Macayeal_1993,Morlighem_2010,Morlighem_2013}. Employing such a minimisation yields estimates of glaciological quantities of interest, but not with accompanying full (posterior) probability distributions with which to characterise uncertainty. A contrasting advantage of using Bayesian methodology for such problems is the provision  of uncertainties on spatially-varying fields (such as the basal friction coefficient) via a posterior probability distribution over all parameters of interest. This comes at a cost, as Bayesian methods require the specification of a complex probabilistic model, and generally involve computational challenges, which have been grappled with in recent literature.

We now give a few (non-exhaustive) examples of statistical problems in glaciology that we will highlight in this chapter:
\begin{itemize}
\item Inferring and predicting ice thickness or ice presence/absence.
\item Inferring subglacial topography (i.e., the geometry of the surface on which a glacier is situated).
\item Inferring surface velocity fields.
\item Inferring basal sliding fields.
\item Inferring spatial fields of net mass gains or losses (e.g., surface mass balance).
\item Inferring the contribution of an ice sheet or glacier to sea-level rise.
\end{itemize}

All of these problems have been addressed on one or more occasions through the use of formal statistical methodology, and the purpose of this review chapter is to highlight some of the works employing such methodology. We will give specific focus to works that are couched in a Bayesian framework, which is particularly useful in glaciology where one needs to combine information from multiple sources in a single model. As we will see, many of these works formulate Bayesian models that are hierarchical, where parameters, geophysical processes, and data processes are modelled separately. An additional advantage of using the Bayesian formalism is that it provides a conditional distribution over unknown quantities given observations; this \emph{posterior distribution} combines information from the data and the \emph{prior distribution}, and yields estimation of unknown quantities as well as uncertainty quantification in light of all the available information.

The overarching aim of this review chapter is to provide an accessible introduction of Bayesian and Bayesian hierarchical modelling approaches in glaciology for earth scientists, as well as a glaciology primer for statisticians. The chapter is structured as follows. In Section 2, we begin with a broad summary of literature that has employed Bayesian statistical modelling (not necessarily hierarchical) in a glaciology context, starting from the late 2000s. The overview emphasises the scientific topics being addressed, the statistical modelling employed, and the conclusions or key take-away messages of the reviewed works. Sections 3 and 4 then give a detailed summary of two studies involving Bayesian hierarchical modelling in the context of glaciology \citep{gopalan2019spatio, zammit2015multivariate}. These sections give more details on the modelling and inferential strategies employed, and clearly demonstrate the role Bayesian statistical methodology has to play in glaciology. Section 5 concludes and discusses current initiatives and research directions at the intersection of Bayesian modelling and glaciology. %

\section{A Synposis of Bayesian Modelling and Inference in Glaciology}

In the current section we review work that has applied Bayesian methods to problems in glaciology in the last two decades. Our aim is to provide the reader with a broad overview of what work has been done at the intersection of Bayesian inference and glaciology, with a view to how future work can build on what has been accomplished to date. We do not focus on the technical details, but rather convey the core research goals of the individual works, their Bayesian modelling strategies, and the statistical computing tools they employ. In contrast, Sections~\ref{sec:SMB} and \ref{sec:RATES} contain an in-depth treatment of two works that incorporate Bayesian statistics in addressing important questions in glaciology. Section~\ref{sec:SMB} tackles the problem of predicting surface mass balance across an Icelandic mountain glacier, while Section~\ref{sec:RATES} details an approach to estimate the Antarctic contribution to sea-level rise. The reader who is interested in these more detailed examples, rather than a broad review, can proceed directly to Section~\ref{sec:SMB}. Appendix A contains details of the governing ice flow equations, referred to in the chapter, that are often used in glaciology.

In this section we categorise papers according to the types of models commonly employed in major Bayesian glaciological modelling studies: Gaussian-Gaussian Bayesian models (Section 2.1), Bayesian hierarchical models (Section 2.2), and models for Bayesian calibration (Section 2.3). 

\subsection{Gaussian-Gaussian Models}
A ubiquitous model that appears in many of the glaciology papers employing Bayesian methods and models is the Gaussian-Gaussian (Gau-Gau) model. 
Assume that the glaciological data of interest is available in a vector $\boldsymbol{x} \in \mathbb{R}^N$, which in turn is related to a number of physical parameters in the vector $\boldsymbol{\theta} \in \mathbb{R}^M$ via some function $f(\cdot)$. The  Gau-Gau model is defined through the distributions:
\begin{eqnarray}\label{eq:Gau1}
p(\boldsymbol{x}\mid\boldsymbol{\theta}) &=& \frac{1}{\sqrt{(2\pi)^N|\boldsymbol{\mathrm{\Sigma}}_l|}}\exp{\left[-\frac{1}{2}(\boldsymbol{x}-f(\boldsymbol{\theta}))^{\mathrm{T}}\boldsymbol{\mathrm{\Sigma}}_l^{-1}(\boldsymbol{x}-f(\boldsymbol{\theta}))\right]},
\end{eqnarray}
and 
\begin{eqnarray}\label{eq:Gau2}
p(\boldsymbol{\theta}) &=& \frac{1}{\sqrt{(2\pi)^M|\boldsymbol{\mathrm{\Sigma}}_p|}}\exp{\left[-\frac{1}{2}(\boldsymbol{\theta}-\boldsymbol{\mu})^{\mathrm{T}}\boldsymbol{\mathrm{\Sigma}}_p^{-1}(\boldsymbol{\theta}-\boldsymbol{\mu})\right]}.
\end{eqnarray}
In \eqref{eq:Gau1} and \eqref{eq:Gau2}, the mean and covariance matrix associated with the  distribution $p(\boldsymbol{x} \mid \boldsymbol{\theta})$ are given by $f(\boldsymbol{\theta})$ and $\boldsymbol{\mathrm{\Sigma}}_l$, respectively. Further, the mean and covariance matrix associated with the prior distribution, $p(\boldsymbol{\theta})$, are given by $\boldsymbol{\mu}$ and $\boldsymbol{\mathrm{\Sigma}}_p$, respectively. We often refer to \eqref{eq:Gau1} and \eqref{eq:Gau2} as the \emph{data} and \emph{process} model, respectively. Furthermore, when (1) is viewed as a function of $\boldsymbol{\theta}$, we refer to it as the \textit{likelihood function}.

While the Gau-Gau model leads to several computational simplifications, there are two complexities that drive much of the research with these models. The first issue concerns that of inverting and finding the determinant of large covariance matrices, which generally has cubic computational complexity. The second issue concerns the function $f(\cdot)$, which determines the conditional mean of the data, $\boldsymbol{x}$. The function $f(\cdot)$ is generally nonlinear, and therefore the posterior distribution over $\boldsymbol{\theta}$ is generally non-Gaussian. As we shall see, many of the papers in the glaciology literature attempt to circumvent these issues with a variety of computational strategies. 

One of the earliest approaches employing Bayesian methods with a Gau-Gau model in a glaciological context is that of \citet{tc-3-265-2009}, who used a Bayesian approach for inferring basal topography and basal sliding of ice streams from observations of surface topography and velocity, and illustrated their methods on simulated data examples. They used Gaussian process priors with squared-exponential kernels for modelling the basal sliding parameters and basal topography (process model), and a physics-based flow forward model, approximated using finite elements, to relate basal conditions to observed surface topography and velocity (data model). A Gauss--Newton iterative optimisation method was then used to compute maximum a posteriori (MAP) estimates of basal sliding and topography. Each iteration required the computation of the gradient of the forward model with respect to the unknown parameters characterising the processes; this was accomplished by using an analytical linear approximation of the forward model. Synthetic tests were used to show that the MAP estimates were often very close to the simulated (true) bedrock velocity and basal sliding profiles. \citet{Pralong} later applied this methodology to infer basal topography and sliding from observations of the Rutford Ice Stream in West Antarctica.

The work of \citet{tc-3-265-2009} was relatively small scale. The increased availability and resolution of remote-sensing instruments, coupled with the need to analyze the behaviour of entire ice sheets, led to work pioneering large-scale tractability for inference with Gau-Gau models. Developments were especially needed for the analysis of entire ice sheets. The papers of \citet{petra2014computational} and \citet{Ghattas} are among the first to address this issue in a glaciological context. In particular, their focus was on Bayesian inference of large, spatial basal sliding fields at glaciers given surface velocity measurements. In both cases, the process and the observations were linked via a nonlinear function motivated by Stokes' ice flow equations, which were implemented using Taylor--Hood finite elements. The computational challenge addressed primarily in \citet{petra2014computational} was how to efficiently sample from the posterior distribution when the dimensionality of the field is high (as in the case of  the Antarctic ice sheets). Their solution was to use a Gaussian (Laplace) approximation \citep[e.g.,][Chapter 13]{gelman2013bayesian} of the posterior distribution over the basal sliding field, where the mean is set to the MAP estimate, and the precision matrix is set to the Hessian of the negative log of the posterior distribution, evaluated at the MAP estimate. For computational tractability, a low-rank approximation was used for the covariance matrix. \citet{Ghattas} used a strategy similar to that developed by \citet{petra2014computational} to obtain a posterior distribution over the basal sliding field in Antarctica, which was then used to obtain a prediction interval for mass loss in Antarctica. 

A Gau-Gau model was also used by \citet{minchew2015early} for inferring glacier surface velocities, which provide valuable information regarding the dynamics of a glacier. In particular, the goal of \citet{minchew2015early} was to infer velocity fields at Icelandic glaciers (Langj\"{o}kull and Hofsj\"{o}kull) solely using repeat-pass interferometric synthetic aperture radar (InSAR) data collected during early June 2012. In the Gau-Gau model of \citet{minchew2015early}, the function $f(\cdot)$ (see Equation (1)) is assumed to be a linear function, and so the posterior distribution is Gaussian with an analytically exact mean and covariance function \citep[e.g.,][Chapter 3]{tarantola2005inverse}. The directions of the inferred horizontal velocities were in the direction of steepest descent, which is consistent with physics models of flow at shallow glaciers (such as those of Iceland). Visualisations of posterior spatial variation in horizontal velocity during the early-melt season (i.e., June) across Langj\"{o}kull and Hofsj\"{o}kull revealed magnitudes of horizontal velocities that differ considerably from those obtained from a physical model that ignores basal sliding velocity. Their results suggest that basal sliding is an important factor to consider when determining the velocities of the Icelandic glaciers Langj\"{o}kull and Hofsj\"{o}kull.

\citet{Brinkerhoff} employed a Gau-Gau modelling approach to estimate subglacial topography given surface mass balance, surface elevation change, and surface velocity, while employing a mass-conservation model for the rate of change of glacial thickness. Gaussian process priors were used for all processes, employing either an exponential or squared-exponential covariance function. Metropolis-Hastings was used for posterior inference via PyMC \citep{patil2010pymc}, and the method was applied to Storglaci\"{a}ren, a 3 km long glacier in northwestern Sweden. The bed topography of  Storglaci\"{a}ren is well-known \citep{Brinkerhoff} and the glacier therefore makes an ideal test for bed topography recovery approaches. \citet{Brinkerhoff} were thus able to show that the estimate (specifically, the MAP estimate) from their model was reasonably close to the true bed topography. Another glacier was also used as a test case - Jakobshavn Isbræ, on the Greenland ice sheet. While the bed topography is not as well known for this glacier, the obtained 95\% credibility intervals  contained bed topography estimates reported elsewhere, although the obtained MAP estimate differed considerably from these other estimates. 

\subsection{Bayesian Hierarchical Models} \label{sec:BHM}
The Gau-Gau model of Section 2.1 is a special case of the more general Bayesian hierarchical model (BHM), which is used to denote models that are constructed via the specification of multiple conditional distributions. To the best of our knowledge, the first work that utilised a Bayesian hierarchical model with more than two levels in a glaciological context is that of \citet{ice_stream}. In this work, the model is couched within a physical-statistical modelling framework \citep{Berliner,cressie2011statistics}, and it is used to infer three processes: glacial surface velocity, basal topography, and surface topography, from noisy and incomplete data. The physical-statistical BHM they used is broken up into three models: a data model, a process model, and a parameter model:

\begin{itemize}
\item The \textit{data model} specifies a statistical model for the data conditional on the physical processes and parameters. 
\item The \textit{process model} specifies a statistical model for the physical processes of interest conditional on one or more parameters.
\item The \textit{parameter model} specifies prior distributions for statistical and/or physical parameters on which inference is made.
\end{itemize}

In \citet{ice_stream}, the aim was to infer the three latent processes from data of surface elevation, basal topography, and surface velocity. A physical argument was used \citep{vanderveen} to relate surface velocity (one of the processes of interest) to stress terms that are in turn dependent on glacial thickness (i.e., the difference between surface elevation and the glacier bed elevation, which are additional physical quantities of interest). Although Gaussianity was assumed at the data and process levels, the posterior distributions over the processes were not Gaussian due to nonlinearity in the conditional mean in the data model. The model was used to analyze data from the Northeast Greenland Ice Stream. Posterior distributions over driving stress, basal sliding velocity, and basal topography were sampled from using a Gibbs sampler and importance-sampling Monte Carlo. 

A BHM that is considerably higher dimensional makes an appearance in \citet{zammit2015multivariate} for assessing the contribution of Antarctica to sea-level rise. Because of the scale of the problem, computational techniques involving sparsity of large precision (i.e., inverse covariance) matrices and a parallel Gibbs sampler were used. A detailed account of this work is given in Section \ref{sec:RATES}.

Several recent papers have used BHMs for inferring unknown parameters and predicting spatially-varying glacial thickness. The main objective of \citet{https://doi.org/10.1002/env.2460} was to infer ice thickness and some parameters of glaciological importance at Thwaites Glacier in West Antarctica. The physics model used was a one-dimensional flowline model that relates the net flux of ice to the accumulation of snow and melting of ice. A shallow ice approximation was used to relate surface velocity to glacial thickness. Surface velocity, slope, ice accumulation, thinning, flow width, and an ice rheological constant are parameters that are input into a glacier dynamics model, which in turn outputs ice thickness along the flowline. Gaussian priors were used for parameters that were treated as unknown, and the likelihood function was Gaussian. A Metropolis--Hastings algorithm was used for posterior inference. 

\citet{gopalan2018bayesian} presented a Bayesian hierarchical approach for modelling the time evolution of glacier thickness based on a shallow ice approximation. As in \citet{ice_stream}, the Bayesian hierarchical model was also couched in a  physical-statistical framework. The process model used was a novel numerical two-dimensional partial differential equation (PDE) solver for determining glacial thickness through time, as well as an error-correcting process that follows a multivariate (Gaussian) random walk. The model was tested on simulated data based on analytical solutions to the shallow ice approximation for ice flow \citep{Bueler}. Inference of a physical parameter (a rheological constant) and predictions of future glacial thickness across the glacier appear biased, although credibility intervals capture the true (simulated) values.

The model was further developed by \citet{Gopalan:2019aa}, with a particular focus on computational efficiency, an issue that must be often addressed in large-scale glaciological applications. Specifically, the use of surrogate process models  constructed via first-order emulators \citep{hooten2011assessing}, parallelisation of an approximation to the log-likelihood, and the use of sparse matrix algebra routines were shown to help alleviate computational difficulties encountered when making inference with large Bayesian hierarchical glaciology models. Additionally, a multivariate random walk assumption in the process model was further developed to include higher-order terms, and it was examined in the context of the shallow ice approximation numerical solver of \citet{gopalan2018bayesian}. This random walk is closely related to the notion of model discrepancy, further discussed in Section 2.3.  

A related contribution to the literature on BHMs in cryospheric science is that of \citet{10.1214/20-BA1209}, which addresses the modelling and prediction of sea ice presence or absence. The data model is a Bernoulli distribution with a temporally and spatially varying probability parameter. The process model characterises the log-odds ratios, derived from the time-varying probabilities, as a linear combination of covariates and spatial basis functions. The coefficients of these basis functions evolve through time according to a vector auto-regressive process of order one. Sampling from the posterior distribution was done using a Metropolis-within-Gibbs sampler. The approach was used to model Arctic Sea ice absence/presence from 1997--2016 using sea ice extent data from the National Oceanic and Atmospheric Administration (NOAA).  

Some of the very recent literature using Bayesian hierarchical models in glaciology has involved non-Gaussian assumptions at either the process, prior, or data levels. For instance, \citet{brinkerhoff_aschwanden_fahnestock_2021} used Bayesian methods for inferring subglacial hydrological and sliding law parameters given a field of time-averaged surface velocity measurements, with application to southwestern Greenland. A physical model was constructed based on an approximate hydrostatic solution to Stokes’ equations coupled with a hydrological model. A surrogate model/emulator was used to reduce the computational complexity of running the full model numerically. Specifically, neural networks (fit using \emph{PyTorch}) were chosen to learn the coefficients of basis vectors as a function of input parameters. The basis vectors were derived from a principal components analysis applied to an ensemble of computer simulator runs. The primary benefit of using the neural network emulator was to reduce the time taken to compute an approximate numerical solution to the surface velocity field. The data model used was multivariate normal, and a beta prior distribution was used for the parameters. Posterior computation was achieved with the manifold Metropolis-adjusted Langevin algorithm, a Markov chain Monte Carlo (MCMC) method that efficiently explores complex posterior distributions \citep{https://doi.org/10.1111/j.1467-9868.2010.00765.x}.

\citet{10.3389/feart.2021.610069} developed a Bayesian approach for inferring a rheological parameter and a basal sliding parameter field at an Icelandic glacier (Langj\"{o}kull). The model relies on the shallow ice approximation for relating physical parameters of interest to surface velocities. Two data models were considered -- one based on the $t$-distribution, and one based on the Gaussian distribution. A truncated normal prior was used for the rheological parameter (termed ice softness), whereas a Gaussian process prior was used for the log of the basal sliding field. Sampling from the posterior distribution was done using a Gibbs sampler with an elliptical slice sampling step \citep{murray2010elliptical} for the basal sliding parameter field. It was found that the inferred rheological parameter was similar to that obtained in other studies, and that the inferred spatial variation in sliding velocity and deformation velocity was generally consistent with that of \citet{minchew2015early}. Residual analysis suggested that a Gaussian data model yields a worse fit than a $t$ data model, although both data models yielded similar posterior distributions over the rheological parameter. This work also clearly demonstrates the utility of uncertainty quantification when inferring important glaciological parameters via a Bayesian inference approach. 

\citet{brinkerhoff_aschwanden_fahnestock_2021} developed a Bayesian approach to jointly estimate a rheological field and a basal sliding parameter field. The approach is similar to that of \citet{10.3389/feart.2021.610069}, except that the rheological parameter is also allowed to vary spatially. The scale of the problem addressed is also larger since the methodology is applied to a large marine-terminating glacier in Greenland. To alleviate computational issues,  a variational inference scheme \citep{doi:10.1080/01621459.2017.1285773} was used instead of MCMC for evaluation of an approximate posterior distribution. A rank-reduction technique, inspired by \citet{Solin:2020aa}, was also used to facilitate computations with large covariance matrices. In contrast to many previous works, \citet{brinkerhoff_aschwanden_fahnestock_2021} used a log-normal distribution for the data model. Moreover, a Gaussian process prior with a squared-exponential kernel was used for both the rheological and basal sliding fields. This paper breaks ground with the use of variational inference instead of MCMC; variational inference has been seeing increased used in recent years in geophysical applications where computational efficiency needs to be addressed.

\subsection{Bayesian Calibration of Physical Models}
Another area of research in glaciology that employs Bayesian methods stems from the computer-model calibration literature \citep{doi:10.1111/1467-9868.00294}. Here, the aim is to tune, or calibrate, numerical models from data in a Bayesian framework. An important consideration in Bayesian calibration is the characterisation of numerical-model discrepancy. Briefly, model discrepancy is a model term used to capture the difference between the output of a computer model (usually the output of the computer model using the \textit{best fitting} parameter value) and the true value of the process being modelled. Model discrepancy has been shown to be necessary in order to obtain realistic inferences of numerical-model parameters \citep{doi:10.1111/1467-9868.00294, discrepancy}.

An early contribution applying ideas from Bayesian calibration to ice-sheet models comes from \citet{mcneall2013potential}, in which the authors developed a Gaussian process emulator of an ice-sheet simulator dubbed Glimmer \citep{rutt2009glimmer} and introduced a method to quantify the extent to which observational data can constrain input parameters (and, consequently, simulator output). In turn, this method may be used to inform the design of observation strategies at ice sheets, as to collect data in order to maximise the extent to which parameters are constrained.

\citet{Chang2016} used a model discrepancy term when developing a Bayesian approach  to calibrate a numerical model that gives a binary output: ice presence or absence. Their method was used to obtain posterior distributions for a variety of physically relevant parameters (including calving factor, basal sliding coefficient, and asthenospheric relaxation $e$-folding time), as well as to make projections in ice volume changes, and consequently predictions of sea-level rise, using the PSU3D-ICE model \citep{pollard2009modelling} and data on the Amundsen Sea Embayment in West Antarctica.  In addition to model discrepancy, \citet{Chang2016} also included an emulator/surrogate model in order to reduce the computational time needed when working with a computationally expensive computer model, much in the spirit of \citet{higdon2008computer}. Emulators are designed to mimic the output of the computer model, but are much less computationally intensive. Inference was made in two steps: First, an emulator for the computer model was constructed using Gaussian processes. Then, the numerical model was calibrated using the emulator and a model discrepancy term, also modelled as a Gaussian process. The work of \citet{Chang2016} and that of \citet{mcneall2013potential} are among the few that use discrepancy models in a glaciological context.

\citet{10.1214/19-AOAS1305} revisited the problem of calibrating Antarctic ice sheet computer models with the goal of improving forecasts of sea-level rise due to ice loss. In contrast to previously summarised approaches at Antarctica, \cite{10.1214/19-AOAS1305} used a parallelisable particle-sampling algorithm for Bayesian computation, dubbed adaptive particle sampling, for reducing computation times. As in \citet{Chang2016}, the PSU3D-ICE model of ice flow was used, though data from the Pliocene era were incorporated into the approach as well. 

In this section we have outlined some major works involving Bayesian statistics and glaciology that fall into three categories: Gau-Gau models, general Bayesian hierarchical models, and Bayesian calibration models. The list of works reviewed is not exhaustive; we also refer the reader to the following papers that contain core elements of Bayesian statistics and cryospheric science: \citet{10.1371/journal.pone.0170052, conrad2018parallel,  nature, Guan, irarrazaval2019bayesian,   tc-14-811-2020, werder2020bayesian, tc-15-1731-2021,10.1214/20-AOAS1405}. Next, in Sections~\ref{sec:SMB} and~\ref{sec:RATES}, we take a detailed look at two case studies that exhibit the utility of Bayesian methods in glaciology.

\section{Spatial Prediction of Langj\"{o}kull Surface Mass Balance}\label{sec:SMB}

We now illustrate an example of latent Gaussian modelling in glaciology with the problem of prediction of surface mass balance (SMB) over Langj\"{o}kull, an Icelandic mountain glacier, following Chapter 6 of the PhD thesis of \citet{gopalan2019spatio}. SMB is the temporal rate of change of mass at the surface of a glacier due to factors such as snow accumulation, ice melt, and snow drift. This quantity is usually an integral part of dynamical equations that involve the temporal derivative of glacier thickness, since a negative SMB contributes to a decrease in thickness, while a positive SMB contributes to an increase in thickness (see governing equations in Appendix A). SMB is usually observed at only a few locations across a glacier, and statistical methods are needed to provide probabilistic predictions at spatial locations away from the measurement sites. 

Glaciologists at the University of Iceland Institute of Earth Sciences (UI-IES) take measurements of SMB twice a year (late spring and fall) across the surface of Langj\"{o}kull, usually at all the 25 sites shown in Figure~\ref{fig:intro}. The measurements taken during late spring are records of winter SMB, whereas the measurements taken during fall are records of summer SMB. The sum of winter and summer SMB yields the net SMB for the year.  

Conventionally, linear models are used in glaciology for predicting SMB at unobserved locations; for instance, \citet{palsson2012mass} show that there is a nearly linear relationship between SMB and elevation except at high elevations, possibly due to snow drift at high elevations. Additionally, a model for precipitation (which is directly linked to SMB) is used by \cite{adhalgeirsdottir2006response} that  is linear in the $x$-coordinate, the $y$-coordinate, and elevation. This is a reasonable model since analyses of stake mass balance measurements from Hofjsj\"{o}kull in Iceland suggest that linearity is an appropriate assumption to make. These works do not, however, model spatial correlations in the residuals, which in turn can lead to suboptimal predictions and overconfident estimates on spatial aggregations of SMB. This is addressed in the model of \cite{gopalan2019spatio}, which we discuss next.

\cite{gopalan2019spatio} used two Bayesian hierarchical models (Section~\ref{sec:BHM}) for making spatial predictions of SMB across Langj\"{o}kull for the summer and winter of every year, respectively, in the recording period 1997--2015. 
The \textit{data model} they used is: 
\begin{align}
Z_{w,t}(\svec) &= SMB_{w,t}(\svec)+\epsilon_{w,t}(\svec), \quad \svec \in \Sset_t,\quad t = 1997,\dots,2015,
\end{align}
where $\Sset_t \subseteq \{\svec_1,\dots,\svec_{25}\}$ is the set of locations at which measurements were made in year $t$, the subscript $w$ refers to winter, and where $\svec_1,\dots,\svec_{25}$ are the 25 measurement locations containing the $x$ and $y$ coordinates from a Lambert projection system (shifted and scaled to lie between 0 to 1). The number of locations in $\Sset_t$ was greater or equal to 22 in the years 1997--2015 across Langj\"{o}kull (earlier years tended to have less measurements than more recent years).  The terms $\epsilon_{w,t}(\svec),\svec \in \Sset_t, t = 1997,\dots,2015,$ are Gaussian white noise terms. The data model used for the summer SMB is analogous to that for winter.

The \textit{process model} model used for SMB is
\begin{align}
SMB_{w,t}(\svec) &= \beta_{0,w,t}+\beta_{1,w,t}s_1+\beta_{2,w,t}s_2+\beta_{3,w,t}z_t(\svec)+U_{w,t}(\svec),~~ \svec \in D. 
\end{align}
As before, the subscript $w$ refers to winter, and the spatial domain of interest, $D$, is the set of $x$-$y$ coordinates in the glacier Langj\"{o}kull. The terms $\beta_{0,w,t},\dots,\beta_{3,w,t}$ are model coefficients corresponding to an intercept, the $x$-coordinate $s_1$, the $y$-coordinate $s_2$, and elevation in metres, $z_t(\cdot)$, in year $t$, respectively, while $U_{w,t}(\cdot), t = 1997,\dots,2015,$ are independent spatial Gaussian processes. Again, the process model for summer is analogous to that for winter. \citet{gopalan2019spatio} fit separate SMB models each year so as to account for year-to-year variation in SMB. The latent Gaussian process $U_{w,t}(\cdot)$ has mean 0 with spatial covariance determined by the Mat\'{e}rn kernel:
\begin{eqnarray}\label{eq:Matern}
        C(\svec_a, \svec_b) &=& \sigma^2\frac{2^{1-\nu}}{\Gamma(\nu)}\left(\frac{\sqrt{8\nu}||\svec_a-\svec_b||}{\rho}\right)^\nu K_\nu\left(\frac{\sqrt{8\nu}||\svec_a-\svec_b||}{\rho}\right),
\end{eqnarray}
for $\svec_a, \svec_b \in D$, where $\sigma$ is the marginal standard deviation,  $\rho$ is the spatial range parameter, $\nu$ is the smoothness parameter, $K_\nu(\cdot)$ is the modified Bessel function of the second kind of order $\nu$, and $\Gamma(\cdot)$ is the gamma function \citep[e.g.,][]{bakka2018spatial}. A Gaussian process with Mat\'{e}rn covariance function \eqref{eq:Matern} is the solution to the  stochastic partial differential equation (SPDE): 
\begin{eqnarray}\label{eq:spde}
(\kappa^2 - \Delta)^ {\alpha/2} (\tau U_{w,t}(\svec)) &=&  {\cal W}(\svec), \quad \svec \in D,
 \end{eqnarray}
 where recall that here $D \subset \mathbb{R}^2$, $\Delta$ is the Laplacian, $\kappa = \sqrt{8\nu}/\rho$, $\alpha = \nu + 1$ (for $d = 2$), ${\cal W}(\cdot)$ is a Gaussian white noise spatial process, and where $\tau>0$ is a function of $\sigma^2, \rho$ and $\nu$ \citep{Lindgren_2015}.

The following specifications were used for the \textit{prior model}. The $\beta$ parameters were all set to have zero mean prior normal distributions. The precision for the intercept term was set to be very small to induce a non-informative prior; a precision of 0.1 was used for the elevation term; and  a precision of 1.0 for the $x$ and $y$ terms. Precisions were selected using a leave-one-out cross-validation procedure. Additionally, we noticed that estimated slopes in year-to-year linear regression fits varied more for elevation than for the $x$ and $y$ coordinates.  Penalised-complexity (PC) priors \citep{simpson2017} were used for the range and scale parameters of the Mat\'{e}rn covariance kernel. 
 
In order to fit the SMB models, \citet{gopalan2019spatio} used the R-INLA software \citep{rue2009approximate}, which employs an approximate Bayesian inference scheme based on nested Laplace approximations for computing approximate posterior distributions over relevant quantities. Additionally, this software implements the scheme of \citet{lindgren2011explicit} in order to approximate the solution to the SPDE in \eqref{eq:spde} as a linear combination of basis functions that are derived from a finite element method (FEM). The FEM leads to the use of sparse precision matrices, with which computations can be done efficiently using sparse matrix algebra routines; see, for example, \citet{golub2012matrix}.

\begin{figure}[t!]
  \centering
    \includegraphics[width=1.0\textwidth]{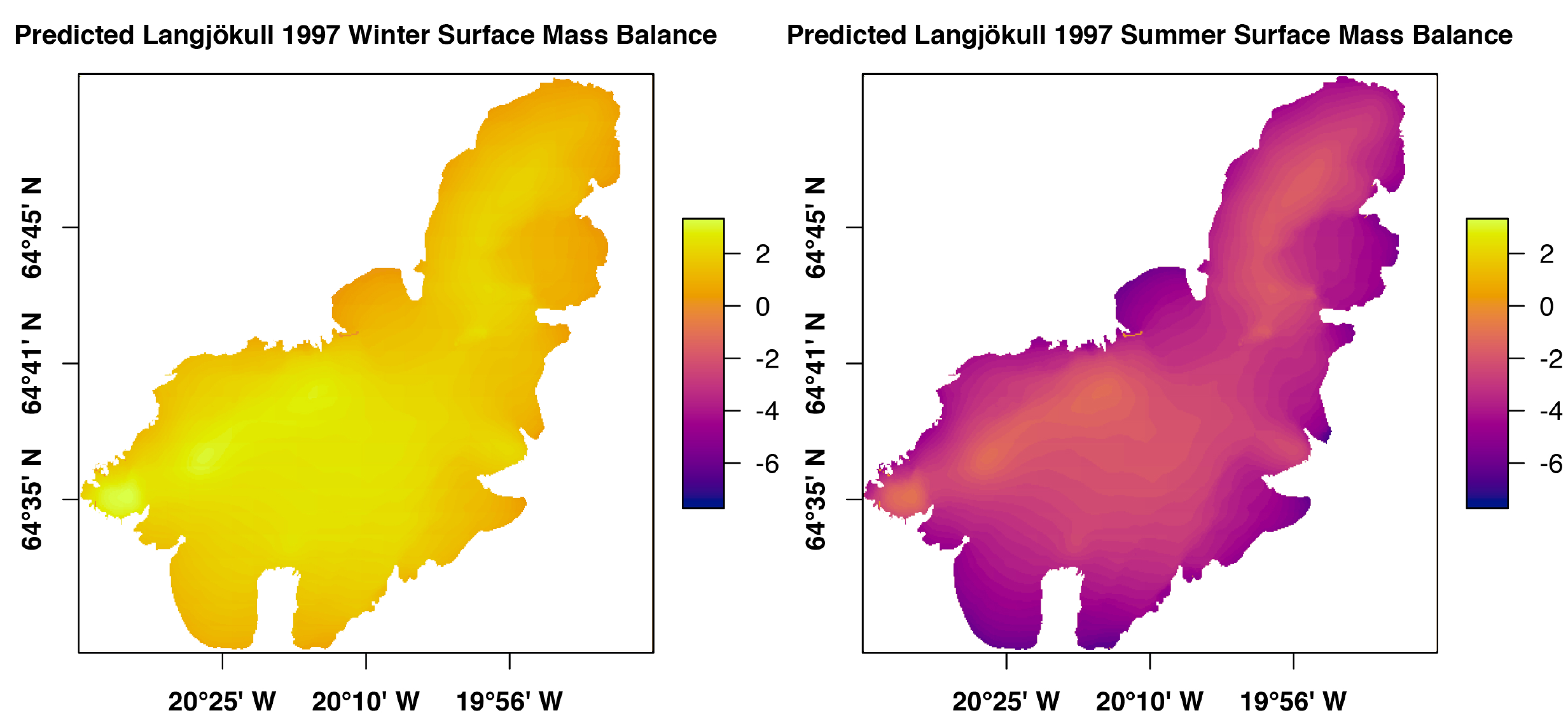}
\caption{Summer and winter SMB predictions, in meters per year water equivalent, across Langj\"{o}kull in the year 1997. Left Panel: Winter SMB predictions. Right panel: Summer SMB predictions.\label{fig:SMB}}
\end{figure}

\begin{figure}[t!]
  \centering
    \includegraphics[width=1.0\textwidth]{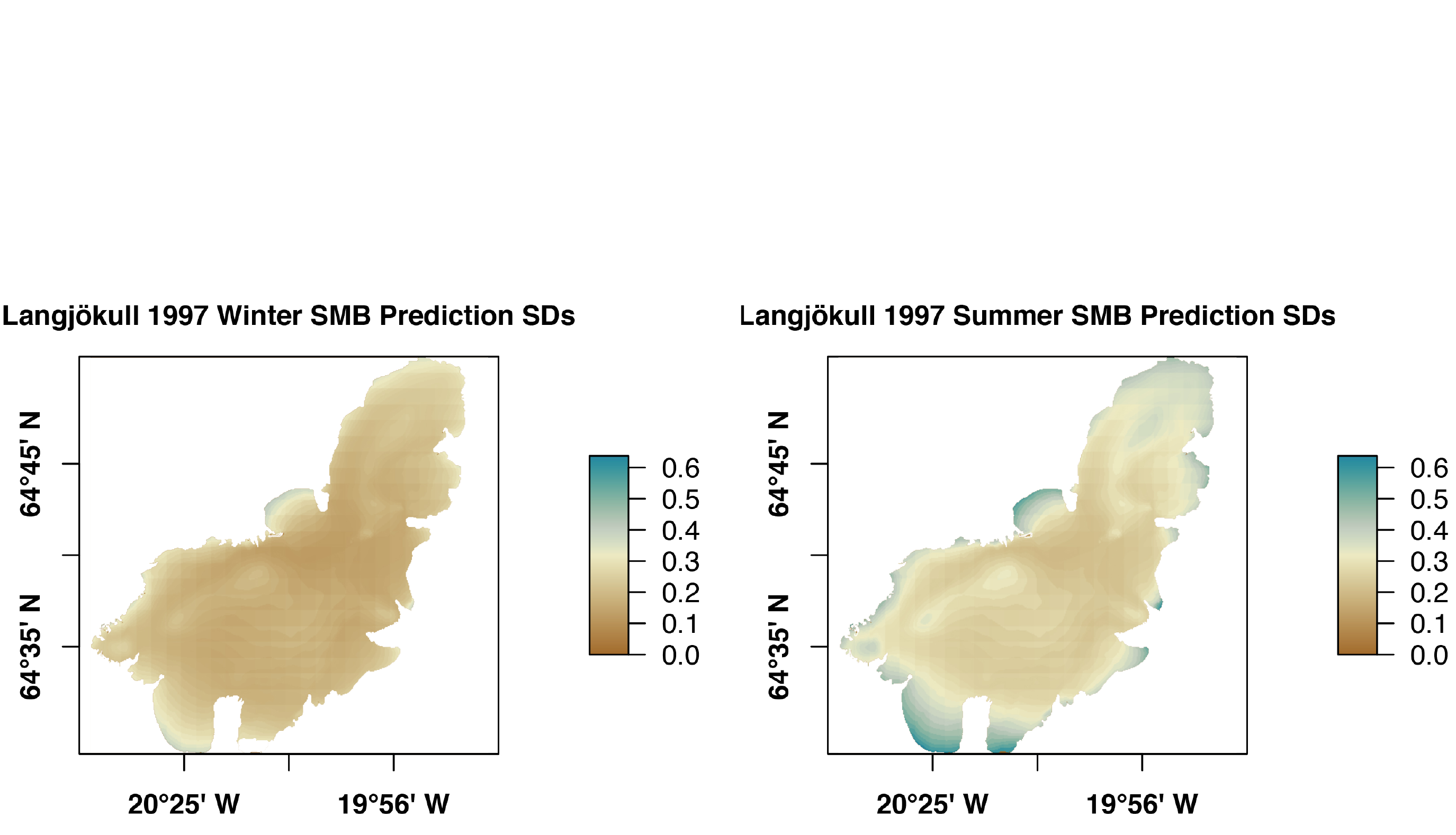}
\caption{Prediction standard deviations (SDs) associated with the predictions of Figure 2, in meters per year water equivalent, across Langj\"{o}kull in the year 1997. Left Panel: Winter SMB prediction SDs. Right panel: Summer SMB prediction SDs.\label{fig:SMBunc}}
\end{figure}

\begin{figure*}[t!]
  \centering
    \includegraphics[width=.7\textwidth]{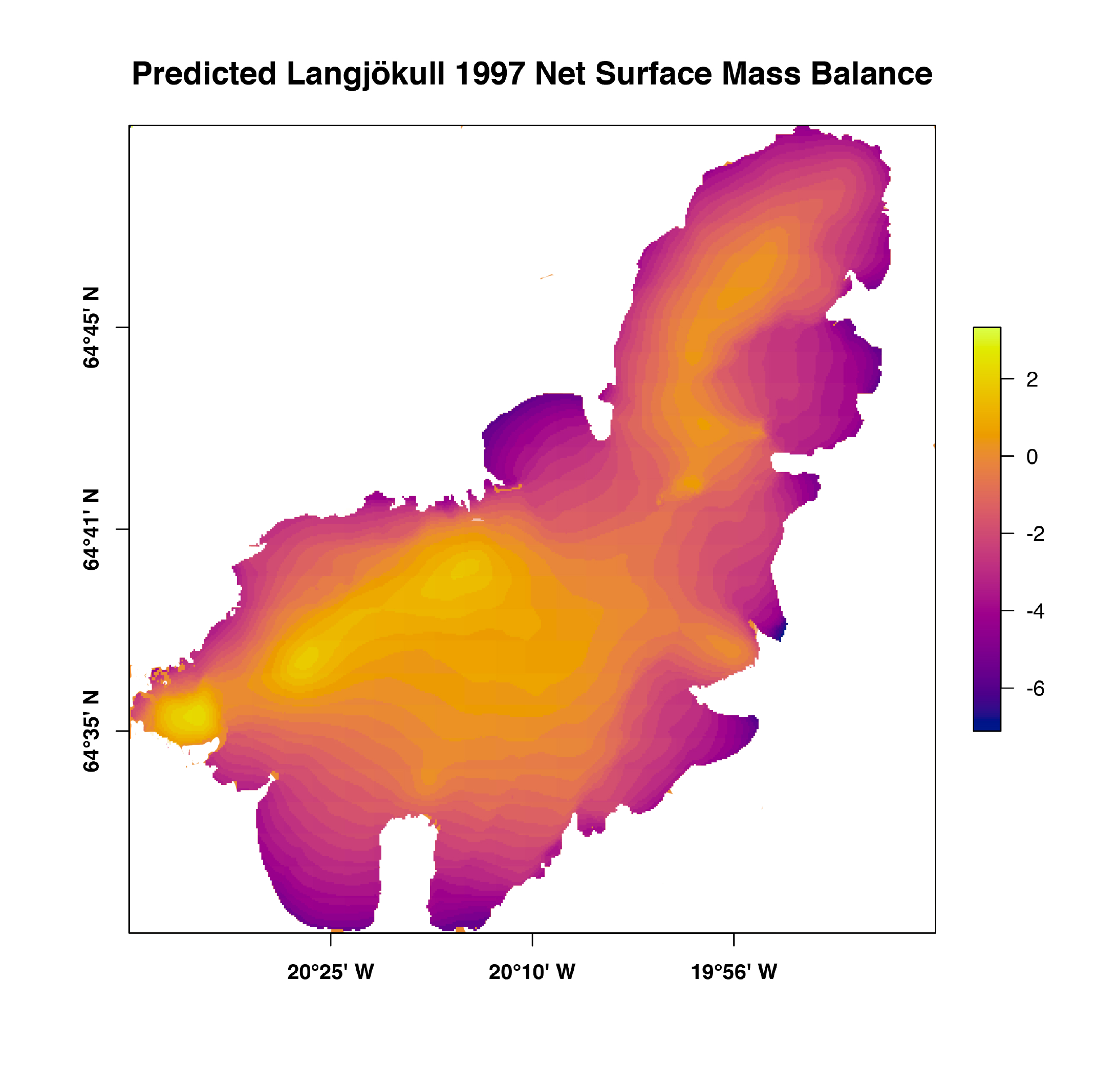}
\caption[Net mass balance predictions during 1997]{Net SMB predictions (sum of summer and winter), in meters per year water equivalent, across Langj\"{o}kull in the year 1997.\label{fig:SMBnet}}
\end{figure*}

\citet{gopalan2019spatio}  fit separate SMB models for summer and winter for each year from 1997 to 2015 and used the fitted models to make predictions for both summer and winter SMB across the glacier at a 100 meter resolution for each year. 
For illustration, the predictions for summer SMB and winter SMB are shown in Figure~\ref{fig:SMB} for the year 1997. Accompanying prediction standard deviations, which are quantitative measures of prediction uncertainty, are shown in Figure~\ref{fig:SMBunc}, while the net SMB is shown in Figure~\ref{fig:SMBnet}. It is clear that, as expected, melting occurs more in the summer than in winter, and that the net SMB tends to be more negative at the perimeter of the glacier, where elevation is lower.  On the other hand, the interior of the glacier tends to see little surface mass loss at the higher elevations.  This general pattern is seen also in later years, although in 2015 the amount of mass loss at the perimeter regions is less than in 1997. These results are consistent with what is expected from baseline physical principles.  

\section{Assessing Antarctica's Contribution to Sea-Level Rise}\label{sec:RATES}

The Antarctic ice sheet is the world's largest potential contributor to sea-level rise, with enough grounded ice to raise the sea level by approximately 60 m \citep{Morlighem_2020}. However, estimating the present day contribution to sea-level rise, let alone the future contribution, is not straightforward: Ice sheets continuously lose mass through basal melting, sublimation, meltwater runoff and flux into the ocean, and gain mass through accumulation of snowfall. When the total mass input matches the total mass output, an ice sheet is said to be in balance, and does not contribute to sea-level rise or lowering. On the other hand, mass loss occurs when there is a shift from the balance state.

Deviations from a balance state are detected through changes in ice surface elevation (altimetry) and changes in mass (gravimetry) across time. For example, a detected loss in ice sheet thickness generally (but not always) corresponds to a loss in ice sheet mass and, hence, a positive contribution to sea-level rise. However, the problem of estimating ice mass loss/gain from data is ill-posed as there are multiple geophysical processes occurring in Antarctica that all contribute to observed change in ice surface elevation and/or change in mass, the most relevant of which are (i) glacio-isostatic adjustment (GIA, which is a solid-earth process), (ii) ice dynamics, (iii) firn compaction, and (iv) a collection of surface processes such as precipitation that affect the mass balance at the surface. Inferring these dynamical processes from observational data is a long-standing problem in glaciology \citep[e.g.,][]{Riva_2009, Gunter_2014}.

\citet{zammit2014resolving} and \citet{zammit2015multivariate} developed a BHM to characterise the multiple processes in Antarctica that contribute to mass balance, which was subsequently used for various studies of the Antarctic ice sheet \citep[e.g.,][]{Schoen_2015, Martin_2016, Chuter_2021}. In the vein of \citet{Berliner} (Section~\ref{sec:BHM}), the BHM is divided into a data model, a process model, and a parameter model. For clarity of exposition, we first describe the process model before proceeding to outline the data and parameter models.

\vspace{0.1in}

{\bf Process model:} The process model of the BHM describes the primary quantities of interest, which in this case are the yearly ice surface elevation changes due to the four geophysical processes listed above. Consider a time index $0,1,\dots,T$, and a spatial domain $D \subset \mathbb{R}^2$. In the BHM,  GIA is modelled as a spatial process $Y_{GIA}(\cdot)$ (since it is practically time-invariant on the time scales of interest) with covariance function \eqref{eq:Matern}, while ice dynamics are modelled as a spatio-temporal regression of the form
\begin{equation}\label{eq:ST_ICE}
Y_{I,t}(\svec) = \xvec_t'\betab(\svec) + w_{I,t}(\svec);\quad \svec \in D, \quad t = 0,1,\dots,T,
\end{equation}
where $Y_{I,t}(\svec)$ is the ice surface elevation that is due to ice dynamics in a time period $t$ (in this case spanning one year) at location $\svec$, $\betab(\svec)$ are spatially varying weights of covariates $\xvec_t' = (1, t)$ (chosen in order to model spatially-varying linear temporal trends), and $w_{I,t}(\cdot)$ is unexplained variation that is modelled as white noise after discretisation (discussed in more detail below). SMB and firn compaction are modelled as conventional autoregressive-1 processes where the error term is spatially correlated. Specifically, for SMB,
\begin{equation}\label{eq:ST}
Y_{S,t}(\svec) = a(\svec)Y_{S,t-1}(\svec) + w_{S,t}(\svec);\quad \svec \in D, \quad t = 1,\dots,T,
\end{equation}
where $Y_{S,t}(\svec)$ is the ice surface elevation change that is due to SMB in a time period $t$, $a(\svec)$ is a spatially-varying autoregressive parameter, and where each  $w_t(\cdot)$ is modelled as an independent spatial process with covariance function \eqref{eq:Matern}. The process model for firn compaction, $Y_{F,t}(\cdot)$, is analgous to that for SMB. These modelling choices can be varied as needed; for example, \citet{Chuter_2021} instead modelled ice surface elevation loss due to ice dynamics as a (highly correlated) autoregressive-1 process in the Antarctic Peninsula, where the assumption of a linear or quadratic temporal trend for all given spatial locations is not realistic.

The processes in the BHM are discretised using the same finite-element scheme described in Section~\ref{sec:SMB}. The triangulation used to construct the elements can vary with the process being modelled; since GIA is smoothly varying, a coarse triangulation suffices for modelling GIA. On the other hand, a fine triangulation is used when modelling ice dynamics, particularly close to the coastline where high horizontal velocities lead to a large variability in ice surface elevation change due to ice dynamics. The finite element scheme is used to establish a finite-dimensional representation of the four processes being considered for each time period $t$. Specifically, for each $i \in \{S, F, I \},t = 0,\dots,T$, the finite-element decomposition of $Y_{i,t}(\cdot)$ yields the model $Y_{i,t}(\cdot) = \phib_i(\cdot)'\etab_{i,t}$, where $\phib_i(\cdot)$ are the finite-element basis functions, and $\etab_{i,t}$ are the coefficients of those basis functions. For GIA, one has that  $Y_{GIA}(\cdot) = \phib_{GIA}(\cdot)'\etab_{GIA}$. The process model is completed by specifying multivariate Gaussian distributions over $\etab_{i,t}, i \in \{S, F, I\},t = 0,\dots,T$, and $\etab_{GIA}$, following the methodology outlined in \citet{lindgren2011explicit}.

\vspace{0.1in}

{\bf Data model:} There are several data sets that could be used to help identify the contributors to observed ice surface elevation change in a glacier or ice sheet. GPS data record changes in bedrock elevation, and are thus useful for estimating GIA. Satellite altimetry records changes in the ice surface elevation and ice shelf thickness (assuming hydrostatic equilibrium). When combined with ice penetrating radar data -- and potentially mass conservation methods \citep[e.g.,][]{Morlighem_2020} -- satellite altimetry can yield ice thickness and bed topography estimates for the grounded ice sheet. Satellite gravimeter instrumentation records gravitational anomalies, which in turn give information on changes in ice-sheet mass as well as mantle flow that occurs during GIA. 

For the data model, one seeks a mapping between the observed quantity and the geophysical process which, recall, could be either spatio-temporal or, in the case of GIA, spatial only (in which case the observations can be viewed as repeated measurements of a spatial-only process). Assume we have $m_t$ observations at time interval $t$, where $t= 0,\dots,T$. In this BHM, these mappings take the general form
\begin{equation}\label{eq:datamodel}
Z_{j,t} = \sum_{i \in \{S, F, I\}}\int_{\Omega_j}f_{i,t}^j(\svec)Y_{i,t}(\svec)\intd \svec +  \int_{\Omega_j}f_{GIA}^j(\svec)Y_{GIA}(\svec)\intd\svec + v_{j,t},
\end{equation}
for  $j = 1,\dots,m_t;~t = 0,\dots,T$, where $Z_{j,t}$ is the $j$th observation at time $t$, $\Omega_j$ is the observation footprint of the $j$th observation, $f_{i,t}^j(\cdot)$ and $f_{GIA}^j(\cdot)$ are instrument-and-process-specific functions which, for example, account for volume-to-mass conversions, and where $v_{j,t}$, which is normally distributed with mean zero and variance $\sigma^2_{j}$, captures both fine-scale process variation and measurement error. For some instruments and processes, $f_{i,t}^j(\cdot) = 0$; for example, $f_{I,t}^j(\cdot) = 0$ if the $j$th datum is from GPS instrumentation, since GPS instruments only measure bedrock elevation change, while $f_{F,t}^j(\cdot) = 0$ if the $j$th datum is from gravimeter instrumentation, since firn compaction is a mass-preserving process. On the other hand $f_{I,t}^j(\svec) = \rho_{I}(\svec)$ if the $j$th datum is from gravimeter instrumentation, where $\rho_I(\svec)$ is the density of ice at $\svec$.

The integrals in \eqref{eq:datamodel} are represented as numerical approximations using the finite-element representations of the processes. Specifically, for the first integral in \eqref{eq:datamodel}, one has that
\begin{equation}
\int_{\Omega_j}f_{i,t}^j(\svec)Y_{i,t}(\svec)\intd \svec \approx \left(\sum_{l = 1}^{n^*_j} f_{i,t}^j(\svec_l)\phib_i(\svec_l)'\Delta_l\right)\etab_{i,t} \equiv (\bvec_{i,t}^{j})'\etab_{i,t},
\end{equation}
where $\svec_l, l = 1\dots,n_j^*,$ denote the centroids of a fine gridding of $\Omega_j$, and $\Delta_l$ denotes the area of each grid cell. The second integral in \eqref{eq:datamodel} is represented in a similar manner. These approximations thus lead to a linear Gaussian relationship between the data and the unknown coefficients $\{\etab_{i,t}\}$ that ultimately represent the processes. Inference proceeds by evaluating the distribution of $\{\etab_{i,t}\}$  conditional on the data $\{Z_{j,t}\}$. When all parameters are assumed known and fixed, this conditional distribution is Gaussian and known. In practice, several parameters also need to be estimated; these are discussed next.

\vspace{0.1in}

{\bf Parameter model:} The BHM requires many parameters to be set or estimated from data. It is reasonable for some of the parameters appearing in the process model, such as temporal and spatial length scales (which themselves vary in different regions of the ice sheet), to be derived from numerical models that describe specific physical processes. In \citet{zammit2015multivariate},  Regional Atmospheric Climate Model \citep[RACMO,][]{Lenaerts_2012} outputs were used to estimate spatio-temporal length scales at various locations on the ice sheet for SMB and firn compaction processes, while the GIA solid-earth model IJ05R2 \citep{Ivins_2013} was used to obtain a spatial length scale ($\rho$ in \eqref{eq:Matern}) for the GIA process. Correlations between SMB and firn compaction, which are also modelled, were estimated from the firn densification model of \citet{Ligtenberg_2011}. One of the main attractions of the process models is that, for several of the processes, they allow one to not place any informative prior belief on the process mean value (often simply called the \emph{prior} in geophysical applications); but solely on the second-order properties (i.e., the covariances) such as length scales, which are relatively well-understood.

There are several other parameters in the BHM that need to be estimated online in conjunction with the weights $\{\etab_{i,t}\}$ that define the latent processes. In the BHM of \citet{zammit2015multivariate}, inference was made over the following parameters that appear in the data and process models:
\begin{itemize}
    \item Parameters that determine the fine-scale variance components $v_{j,t}, j = 1,\dots,m_t; t = 0,\dots,T$ that were, in turn, allowed to vary with characteristics of the surface topography.
    \item Parameters that determine the extent of induced spatial smoothing in gravimetry-derived products.
    \item Multiplicative scaling factors that inflate or deflate the error variances  supplied with the data products, when there is evidence that these are too large or too small.
    \item Parameters used to construct prior distributions for spatially-varying weights $\betab(\cdot)$ in \eqref{eq:ST}. The variances of these weights are generally spatially varying; for example, when modelling ice surface elevation change due to ice dynamics, the variance of the spatially-varying temporal linear trend was constrained to be a monotonically increasing function of horizontal ice velocity.
\end{itemize}

\begin{figure}[t!]
\begin{center}
 	\includegraphics[width=0.6\linewidth]{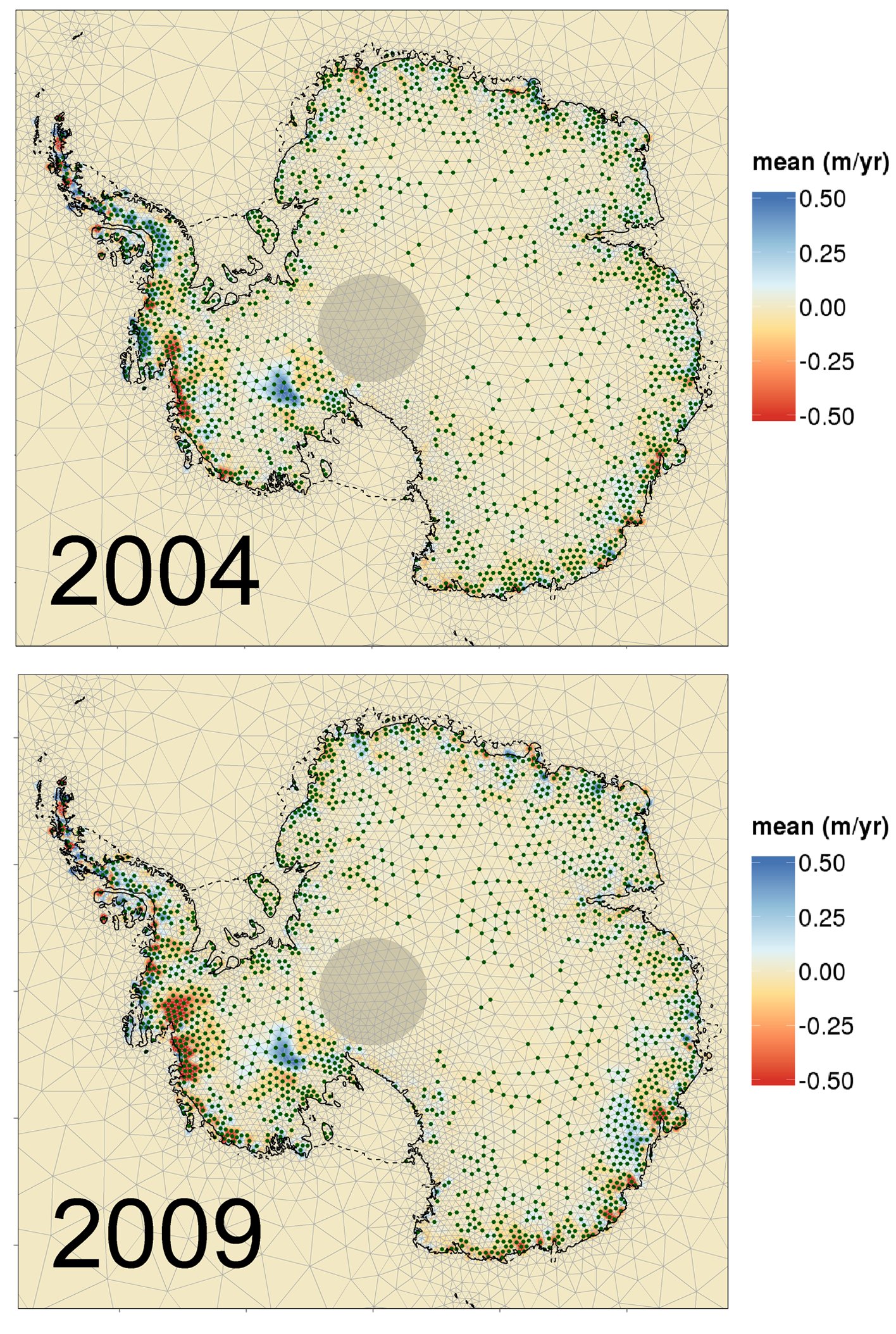}
 	\caption{Posterior mean rate of change of ice surface elevation due to ice dynamics for the years 2004 and 2009. Ice surface elevation changes over ice shelves (enclosed between the dashed and solid lines, which are the coastline and grounding line, respectively) are omitted as these do not contribute to sea-level change. The triangulation (grey lines) is that used to establish the finite-dimensional representation for $Y_{I,t}(\cdot)$ (see \emph{Process model} in Section~\ref{sec:RATES}). Stipples (green dots) denote areas where the posterior mean is significantly different from zero (in this case, where the ratio of the posterior standard deviation to the absolute posterior mean is less than one). Figure adapted from \citet{zammit2015multivariate}.} \label{fig:Results_ICE}
\end{center}
\end{figure}

Inference with the BHM was done using Metropolis-within-Gibbs MCMC, yielding samples from the posterior distributions over all the unknown processes and those parameters that were estimated concurrently with the processes' weights. The authors found broad agreement between their inferences, which are for the years 2003--2009, and numerical-model outputs, and also reported mass balance changes that largely agree, within uncertainty, with those from other approaches that make heavy use of numerical-model output in their analyses. For illustration, Figure~\ref{fig:Results_ICE} is an adaptation of Figure~8 of \citet{zammit2015multivariate}, and depicts the inferred ice surface elevation change in Antarctica that is occurring due to ice dynamics in the years 2004 and 2009.

We conclude this section with a brief discussion on the benefits and drawbacks of employing this fully Bayesian approach for modelling the Antarctic contribution to sea-level rise. The advantages are several: First, the posterior inferences are predominantly data-driven and largely agnostic to the underlying physical drivers. They therefore can also be used to validate output from solid-earth or ice-sheet models. Second, uncertainties are provided on all geophysical quantities, which are interpretable both intra-process and inter-process. For example, a strong anti-correlation between the ice surface elevation contribution from SMB and ice dynamics in the posterior distribution is indicative of the difficulty in separating out the contributors of the observed elevation change; this uncertainty is allowed to vary spatially as well as temporally. Third, it provides a principled way with which to integrate expert knowledge, via the prior distributions placed over the processes and the parameters, with observational data, which can be heterogeneous and have differing support. The limitations are the following: First, it is computationally intensive to make inference with the BHM, and further increasing the spatial and temporal resolution could result in computational difficulties. Second, useful inferences heavily rely on the provision of high-quality data; unrecognised biases or strongly correlated errors in the data products used will adversely affect the predictions and will not be acknowledged in the returned prediction (uncertainty) intervals. Finally, to be used correctly, the framework requires both statistical expertise and glaciological expertise, and is thus only useful for teams with experts in both disciplines. Software that automatically performs this type of analysis is lacking; this is an important angle of statistics for cryospheric science that requires further development. Code for reproducing the results in \citet{zammit2015multivariate} is available from \url{https://sites.google.com/view/rates-antarctica/home}.

\section{Conclusions and Future Directions}

Collecting data from mountain glaciers and ice sheets is a laborious and time-intensive process. For instance, the Antarctic Ice Sheet is the most remote place on earth and experiences some of the harshest conditions in the world. Advances in remote sensing have led to extended spatial coverage of ice surface fields (e.g., elevation and velocity) and ice mass change over recent decades, facilitating advances in our understanding of ice dynamic processes and evolution. However, some data -- including bed topography that is crucial in predicting Antarctica’s contribution to future sea-level rise -- remain sparsely and unevenly sampled, and come at a premium. In such circumstances, the BHM is a powerful framework to maximise the utility of available data, by: (i) allowing one to easily integrate data from multiple sources and with differing support, (ii) allowing one to incorporate prior information and effectively exclude implausible estimates a priori, (iii) allowing one to incorporate dynamical models or simulations into a statistical model, and (iv) providing a means to quantify uncertainty of the inferential target or forecast, potentially informing scientists where to invest next for expensive data collection. 

 Bayesian statistics is a mature field of research, but further investigation and development of several techniques in the field of cryospheric sciences is warranted. For example, the vast majority of the chapter discussed models that are based on underlying Gaussian assumptions. While Gaussianity may be a reasonable assumption for the process (i.e., the latent) model, Gaussianity is often an unrealistic assumption for the data model, particularly in situations where the measurement-error distribution exhibits skewness and heavy tails. When Gaussianity is a valid assumption, there are computational challenges involved that may be addressed using advances in numerics and computational statistics. The reviewed literature has showcased some of these techniques, including the use of surrogate models, the use of sparse matrix algebraic routines, and the use of parallelisable inference schemes. It is clear that developments in Bayesian computation strategies will also benefit the inferential aspects of cryospheric science. 

Despite the benefits of Bayesian reasoning and analysis, there is a paucity of Bayesian methods used by the cryospheric science community. For example, out of the 1744 articles published in the flagship journal \emph{The Cryosphere} between the years 2013 and 2021, only 16 articles contain the word Bayes, or derivatives thereof, in the title, abstract, or keywords, and two of these are co-authored by authors of this chapter. However, recent years have seen a sustained increase in investment in the use of Bayesian methods for the cryospheric sciences. For example the \textit{GlobalMass} project\footnote{https://www.globalmass.eu} that started in the year 2016 is a five-year European Union project where the BHM outlined in Section~\ref{sec:RATES} is applied to other ice sheets and glaciers, as well as other processes that contribute to sea-level rise, while \textit{Securing Antarctica's Environmental Future}\footnote{https://arcsaef.com/} that started in the year 2021 is a seven-year Australian Research Council initiative that will see, amongst other things, Bayesian methods being implemented for data fusion and statistical downscaling to model and generate probabilistic forecasts of changes in environmental conditions and biodiversity in Antarctica. This increased use of Bayesian methods and investment is expected to be sustained for many years to come as data availability and the awareness of the importance of uncertainty quantification increases over time.

\section*{Acknowledgements}

Zammit-Mangion was supported by the Australian Research Council (ARC) Discovery Early Career Research Award (DECRA) DE180100203. McCormack was supported by the ARC DECRA DE210101433. Zammit-Mangion and McCormack were also supported by the ARC Special Research Initiative in Excellence in Antarctic Science (SRIEAS) Grant SR200100005 Securing Antarctica’s Environmental Future. The authors would like to thank Bao Vu for help with editing the manuscript, as well as Haakon Bakka for providing a review of an earlier version of this manuscript.

\section*{Appendix A: Governing Equations}
For ease of exposition, in this appendix  we omit notation that establishes dependence of a variable or parameter on space and time. All variables and parameters should be assumed to be a function of space and time unless otherwise indicated.

Modelling ice sheet flow relies on the classical laws of conservation of momentum, mass, and energy. For incompressible ice flow, conservation of momentum is described by the full Stokes equations:
\begin{align}
    \nabla\cdot\boldsymbol{\sigma} + \rho \boldsymbol{g} &= \boldsymbol{0},\\
    \text{Tr}(\boldsymbol{\dot{\varepsilon}})&=0,    
\end{align}
where $\nabla\cdot\boldsymbol{\sigma}$ is the divergence vector of the stress tensor $\boldsymbol{\sigma}$, $\rho$ is the constant ice density, $\boldsymbol{g}$ is the constant gravitational acceleration, $\boldsymbol{\dot{\varepsilon}}$ is the strain rate tensor and $\text{Tr}$ is the trace operator. The stress and the strain rate are related by the material constitutive relation:
\begin{equation}
    \boldsymbol{\sigma'}=2\eta\boldsymbol{\dot{\varepsilon}},
\end{equation}
where $\boldsymbol{\sigma'}=\boldsymbol{\sigma}+p\boldsymbol{I}$ is the deviatoric stress tensor, $p$ is the pressure, $\boldsymbol{I}$ is the identity matrix, and $\eta$ is the viscosity. Boundary conditions for the mechanical model typically assume a stress-free ice surface, and the specification of a friction or sliding law at the ice-bedrock interface. 

A number of simplifications to the full Stokes equations exist, including the three-dimensional model from Blatter (1995) and Pattyn (2003); the two-dimensional shallow-shelf approximation (MacAyeal, 1989); and the two-dimensional shallow-ice approximation (Hutter, 1983). 

Conservation of mass is described by the mass transport equation:
\begin{equation}
    \frac{\partial H}{\partial t}+\nabla\cdot H\boldsymbol{v} = M_s + M_b,
\end{equation}
where $H$ is the ice thickness, $M_s$ is the surface mass balance, and $M_b$ is the basal mass balance. For regional models of ice flow, the thickness is prescribed at the inflow boundaries, and a free-flux boundary condition is typically applied at the outflow boundary.

Finally, conservation of energy is described by the following equation:
\begin{equation}
    \frac{\partial T}{\partial t} = (\boldsymbol{w}-\boldsymbol{v})\cdot\nabla T+\frac{k}{\rho c}\Delta T+\frac{\Phi}{\rho c}
\end{equation}
where $T$ is the ice temperature, $k$ is the constant thermal conductivity, $c$ the constant heat capacity, and $\Phi$ the heat production term. The boundary conditions typically constitute a Dirichlet boundary condition at the ice surface, a relation for the geothermal and frictional heating at the base of the ice sheet, and a relation for the heat transfer at the ice-ocean interface.

\bibliographystyle{apalike} 

\bibliography{reference_list_glaciology}

\end{document}